%% file: main.tex
\documentclass[conference]{IEEEtran}

\IEEEoverridecommandlockouts
\usepackage{cite}
\usepackage{amsmath,amssymb,amsfonts}
\usepackage{algorithmic}
\usepackage{graphicx}
\usepackage{textcomp}
\usepackage{xcolor}
\renewcommand{\baselinestretch}{0.837}
\usepackage{tikz}
\usepackage{pgfplots}
\usepackage{pgfplotstable}
\usepackage[hmargin=2cm,vmargin=2cm]{geometry}
\pgfplotsset{compat=1.16}
\usetikzlibrary{patterns}
\usepackage{enumerate}
\usepackage[caption=false]{subfig}
\usepackage[capitalise]{cleveref}
\usepackage[export]{adjustbox}

\newcommand*{\affaddr}[1]{#1} % No op here. Customize it for different styles.
\newcommand*{\affmark}[1][*]{\textsuperscript{#1}}
\newcommand*{\email}[1]{\texttt{#1}}

\def\BibTeX{{\rm B\kern-.05em{\sc i\kern-.025em b}\kern-.08em
    T\kern-.1667em\lower.7ex\hbox{E}\kern-.125emX}}
\begin{document}

\title{On BTI Aging Rejuvenation in Memory Address Decoders}
%\thanks{Identify applicable funding agency here. If none, delete this.}
%}
%\iffalse
\author{
Cemil Cem Gürsoy\affmark[1], Daniel Kraak\affmark[2], Foisal Ahmed\affmark[1], Mottaqiallah Taouil\affmark[2],\\ Maksim Jenihhin\affmark[1], and Said Hamdioui\affmark[2]\\
\affaddr{\affmark[1]Department of Computer Systems, Tallinn University of Technology,}\\
 19086 Tallinn, Estonia, \email{\{cemil.gursoy,foisal.ahmed,maksim.jenihhin\}@taltech.ee}\\
\affaddr{\affmark[2]Department of Quantum and Computer Engineering, Delft University of Technology, }\\
2628 CD Delft, The Netherlands,\email{\{D.H.P.Kraak,M.Taouil,S.Hamdioui\}@tudelft.nl}\\
}

\IEEEoverridecommandlockouts \IEEEpubid{\makebox[\columnwidth]{978-1-6654-5707-1/22/\$31.00~\copyright2022 IEEE \hfill} \hspace{\columnsep}\makebox[\columnwidth]{ }}
\maketitle
\IEEEpubidadjcol
\input{0-Abstract}
\input{1-Introduction}
\input{2-Background}
\input{3-Proposed_Mitigation_Methodology}
\input{4-Experimental_Results}

\input{5-Discussion}
\input{6-Conclusion}

\section*{Acknowledgments}
This work was partly supported by the European Union through the European Social Fund in the frames of the ‘‘Information and Communication Technologies (ICT) programme’’ (“ITA-IoIT” topic), by the Estonian Research Council grant PUT PRG1467 “CRASHLESS", and by the RESCUE funded from the European Union’s Horizon 2020 research and innovation program under the Marie Sklodowaska-Curie grant agreement No. 722325.
\bibliographystyle{IEEEtran}

%\bibliography{bibliography}
% Generated by IEEEtran.bst, version: 1.14 (2015/08/26)

\end{document}

%% file: 0-Abstract.tex
\begin{abstract}
Memory designs require timing margins to compensate for aging and fabrication process variations. With technology downscaling, aging mechanisms became more apparent, and larger margins are considered necessary. This, in return, means a larger area requirement and lower performance for the memory. Bias Temperature Instability (BTI) is one of the main contributors to aging, which slows down transistors and ultimately causes permanent faults. In this paper, first, we propose a low-cost aging mitigation scheme, which can be applied to existing hardware to mitigate aging on memory address decoder logic. We mitigate the BTI effect on critical transistors by applying a rejuvenation workload to the memory. Such an auxiliary workload is executed periodically to rejuvenate transistors that are located on critical paths of the address decoder. Second, we analyze workloads' efficiency to optimize the mitigation scheme. Experimental results performed with realistic benchmarks demonstrate several-times lifetime extension with a negligible execution overhead. 
\end{abstract}

\begin{IEEEkeywords}
BTI, aging, rejuvenation, mitigation, memory, address decoder
\end{IEEEkeywords}

%% file: 1-Introduction.tex
\section{Introduction}

The continuous miniaturization of devices has been the main driver of improvements in the semiconductor industry. On the other hand, reliability threats have become more severe due to this trend~\cite{SH13}. Bias Temperature Instability (BTI) is considered to be the main contributor to the time-dependent variability of nanometer-scale devices~\cite{KB06,KKK10}. BTI slows down transistors over time, thus potentially creating reliability issues, such as delay faults.
Traditionally, designers use guard-banding to tolerate this time-dependent variability, i.e., margins are added to the design to ensure a reliable operation.
A downside of this approach is that these margins result in a penalty in area, power, and performance.
Alternatively, designers can embed mitigation schemes that reduce the impact of aging into their design.
In this work, we propose an aging mitigation scheme for the memory address decoder logic.
Memories are a fundamental part and cover a large area of modern Integrated Circuits (ICs). Therefore, they are critical for the overall reliability of a system. In particular, delay faults in the decoder logic contribute to a significant portion of the customer returns~\cite{WN98,AJG04}. Cumulative delay due to BTI-induced aging together with the delay from process variations on decoder logic may cause a wrong address selection during a read or write operation, and, consequently, read or write failures.

%In literature, there are numerous works that analyzes BTI aging effects in memories~\cite{SVK06,IA13,SK13,SK14,IA14,JD15,IA17,JK17,DK19}.
In literature, most work has focused on analyzing and mitigating the impact of aging on the memory cells~\cite{SVK06,IA13,SK13,SK14,IA14,JD15,IA17,JK17,DK19}. These mitigation techniques are mainly based on balancing ones and zeros that are stored in the memory cells, since this reduces the BTI aging impact. There is significantly less work on the peripheral structures of memories. For example, mitigation schemes have been proposed for the Sense Amplifier~\cite{DK17,IA19}.
To the best of our knowledge, there are only two works that target address decoders. 
The authors of~\cite{sw-based} introduce a software-based scheme that mitigates aging by periodically running a rejuvenation workload on top of a user workload. However, it does not provide a method to run the scheme, analyzes simplistic workloads and only targets the Negative BTI (NBTI) aging mechanism. 
The authors of~\cite{hw-based} propose a hardware-based mitigation scheme for address decoders that takes advantage of idle cycles to change the decoder's address input, thereby reducing static BTI stress.
A downside of that approach comes from its hardware overhead (area, power and delay). In addition, it may even lead to a higher aging-induced degradation if the idle period is too long.

%Why existing solutions for logic do not apply for decoders?~\cite{MJ16}
% 8714787 - workload-aware SA BTI aging prediction

In this paper, we propose a low-cost software-based aging mitigation scheme to extend the lifetime of a  memory's address decoder up to several times. The scheme can be applied to existing hardware, and only requires a minor modification to its software. It mitigates aging by running a \emph{rejuvenation} workload \textit{periodically} during the main functional operation of the system. We propose several approaches to generate such workloads. The rejuvenation workload has an opposite effect (in terms of BTI aging) with respect to the functional workload and, thus, it helps transistors on long paths of the decoder to recover from BTI-induced aging. Our experimental results show that it is possible to recover a significant part of BTI aging at a minimal execution overhead. Our contributions in this paper are threefold:
\begin{enumerate}[1.]
    \item It proposes a low-cost aging mitigation scheme to extend the lifetime of a memory's address decoder up to several times.
    \item It proposes a design and workload-aware methodology to generate optimized rejuvenation workload which can be used with the scheme.
    \item It validates  our proposed scheme and proposed rejuvenation workloads using realistic user workloads and two different decoder designs.
\end{enumerate}
   
The rest of the paper is organized as follows. Section II provides background on the BTI mechanism and address decoders. Section III introduces our proposed mitigation methodology. Section IV presents the experimental setup, performed experiments, and the obtained results. Sections V and VI provide a brief discussion and conclusions for the paper.

%% file: 2-Background.tex
\section{Background}

In this section, we briefly introduce the adopted BTI aging mechanism and the address decoders. Then, we explain how BTI influences an address decoder.

\subsection{Bias Temperature Instability model}

% Fig. 1
\begin{figure}[h!]
\includegraphics[width=\linewidth]{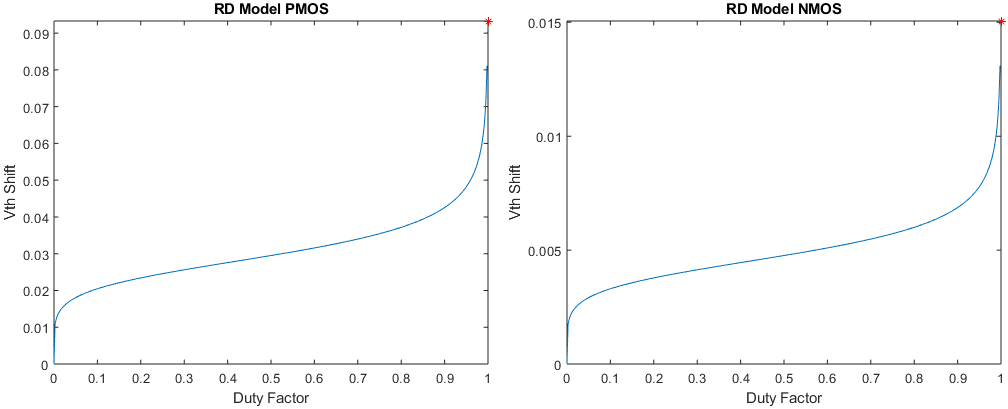}
\centering
\caption{The adopted BTI aging model for PMOS and NMOS transistors.}
\label{fig:RDModel}
\end{figure}

In this study, we rely on the accurate combined model for Negative BTI in PMOS and Positive BTI in NMOS transistors proposed in \cite{rzepa2018comphy} for the 28 nm technology and calibrated for 22 nm Predictive Technology Model (PTM) technology as explained in \cite{sw-based}. The resulting dependency of the BTI-induced threshold voltage (Vth) shift depending on the average duty factor (the probability of a transistor to be on) for NMOS and PMOS is presented in \cref{fig:RDModel}. 

\subsection{Address Decoder}

Address decoders in memories are responsible for accessing the desired cells in the memory cell array. In order to access a particular row and column in the memory cell array, a \emph{wordline decoder} and a \emph{column decoder} are used, respectively. In general the wordline decoder is more critical, since memories typically have more rows than columns. For this reason, we analyze two different 9-to-512 wordline decoder designs in this work, namely a NAND-NOR decoder and an AND-AND decoder.

\begin{figure}[t!]
\centering
\subfloat[\label{fig:decoder_schematic}]{
    \includegraphics[width=0.47\linewidth]{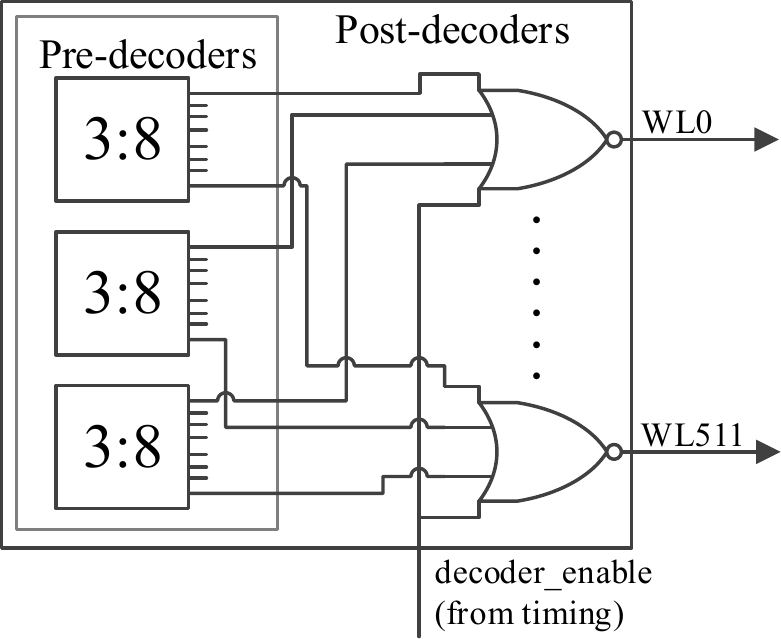}
}%
\subfloat[\label{fig:predecoder}]{
    \includegraphics[width=0.47\linewidth]{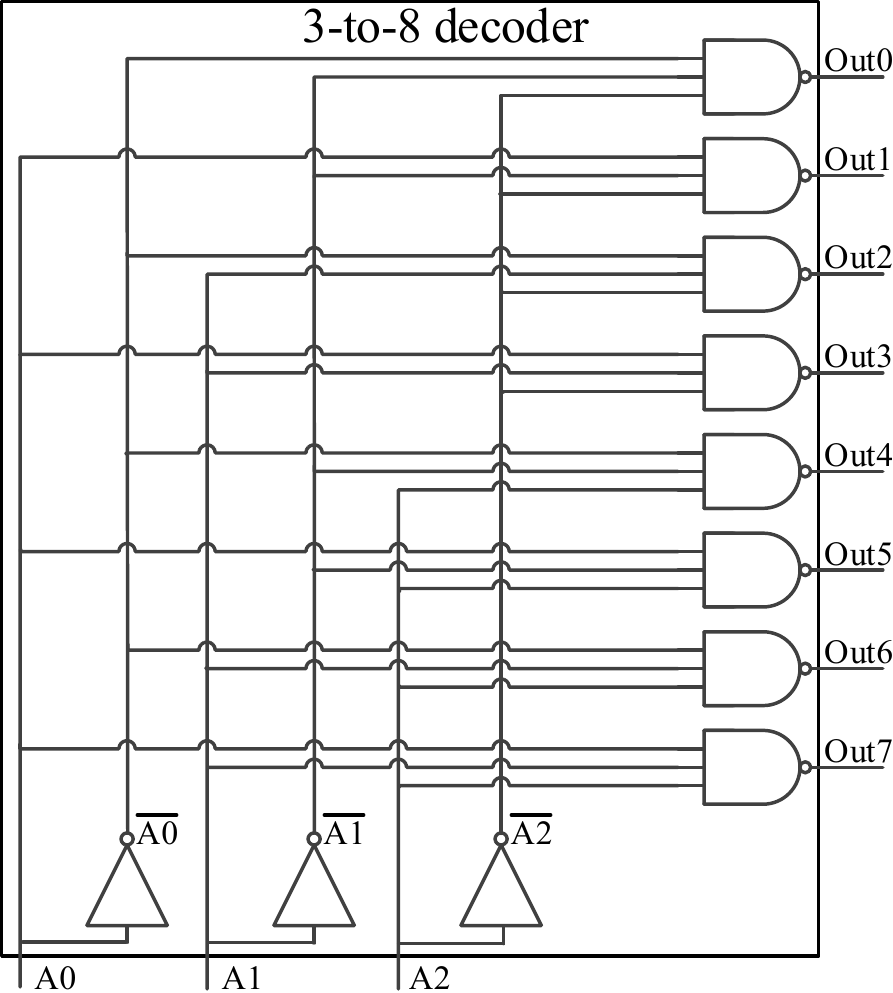}
}%
\hfill
\subfloat[\label{fig:decoder_metric}]{
    \includegraphics[width=0.8\linewidth]{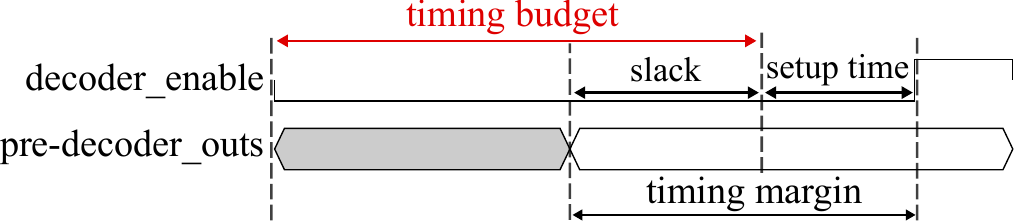}
}%
\caption{(a) Schematic of the wordline decoder and (b) a pre-decoder. (c) The slack metric.}
\label{fig:decoder_schematic_metric}
\end{figure}

\cref{fig:decoder_schematic} shows a simplified schematic of the NAND-NOR decoder. It is a hierarchical design, meaning it consists of a \emph{pre-decoder} and a \emph{post-decoder} stages. The pre-decoder stage is implemented with three 3-to-8 decoders that have the address bits as their input.
%%Fig. 2b shows the schematic of the pre-decoder. As can be seen, it consists of inverters and NAND-gates...
In our first decoder design, the pre-decoder consists of inverters and NAND-gates. Unique combinations of the original and inverted address bits are fed to the inputs of the NAND-gates. This way, each input combination to the decoder results in one of its outputs becoming low. The post-decoder stage consists of 512 post-decoders that are implemented using NOR-gates. Each post-decoder activates one of the wordlines of the memory cell array. 
It has as its inputs a unique combination of the outputs from the pre-decoders.
In addition, it has as input a \textit{decoder\_enable} signal, that is generated by the timing circuit.
Using this signal it is possible to control the duration of the wordline activation.
The AND-AND wordline decoder is implemented similarly as the NAND-NOR decoder. In this case, the pre-decoders consist of AND-gates and inverters, and the post-decoder stage is implemented using AND-gates.

% Fig. 2

\cref{fig:decoder_metric} shows an important reliability metric of the decoder with respect to timing signals, the \emph{slack} time. It is defined as the time between the pre-decoder outputs being ready and the setup time of the post-decoder inputs. Before the \emph{decoder\_enable signal} is activated, the pre-decoder outputs must be ready and stable during \emph{setup time}. If the pre-decoder outputs take too long to settle, the slack becomes negative and the setup time is violated. This may lead to wrong address selection or selection of multiple wordlines and, thus, read or write failures. Increasing timing budget to accommodate more slack time results in more reliable operation, but the memory performance degrades.

\subsection{Aging in Address Decoders}

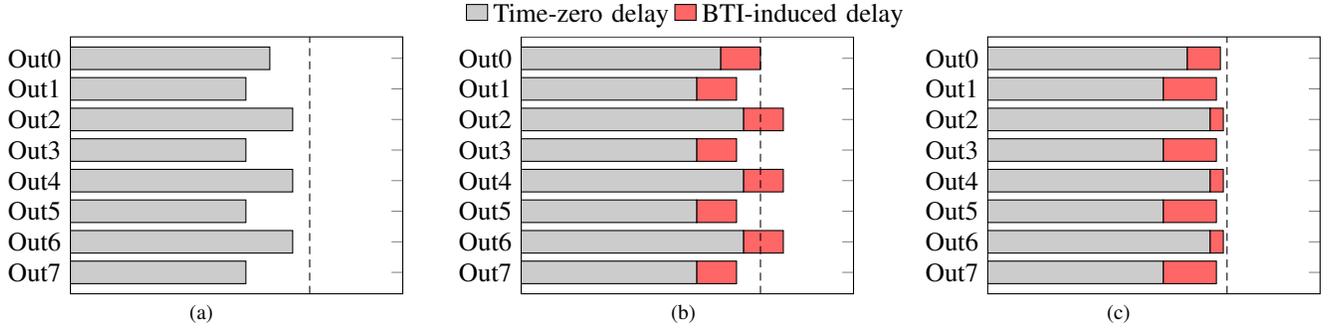
\begin{figure*}[t]
\centering
\pgfplotstableread{
Y Label Nominal
1 Out7 1.320
2 Out6 1.672
3 Out5 1.320
4 Out4 1.672
5 Out3 1.320
6 Out2 1.672
7 Out1 1.320
8 Out0 1.5
}\datatable
\subfloat[\label{fig:path_delays_a}]{%
\begin{tikzpicture}
\begin{axis}[
    %ticks=none,
    xmajorticks=false,
    %axis x line=none,
    %axis y line=left,
    width=6cm,
    height=5cm,
    xbar stacked,
    %xmajorgrids,
    xmin=0,
    xmax=2.5,
    %xtick distance=0.5,
    enlarge y limits={0.1},
    bar width=3mm,
    ytick=data,
    yticklabels from table={\datatable}{Label},
    ]
\draw [black, densely dashed] (1.8,\pgfkeysvalueof{/pgfplots/ymin}) -- (1.8,\pgfkeysvalueof{/pgfplots/ymax});
\addplot [fill=gray!40!white] table [x=Nominal, y=Y] {\datatable};
\end{axis}
\end{tikzpicture}%
}
\hspace{5mm}
\pgfplotstableread{
Y Label Nominal Aged
1 Out7 1.320 0.3
2 Out6 1.672 0.3
3 Out5 1.320 0.3
4 Out4 1.672 0.3
5 Out3 1.320 0.3
6 Out2 1.672 0.3
7 Out1 1.320 0.3
8 Out0 1.5 0.3
}\datatable
\subfloat[\label{fig:path_delays_b}]{%
\begin{tikzpicture}
\begin{axis}[
    %ticks=none,
    xmajorticks=false,
    %axis x line=none,
    %axis y line=left,
    width=6cm,
    height=5cm,
    xbar stacked,
    %xmajorgrids,
    xmin=0,
    xmax=2.5,
    %xtick distance=0.5,
    enlarge y limits={0.1},
    bar width=3mm,
    ytick=data,
    yticklabels from table={\datatable}{Label},
    legend style={draw=none,anchor=south,at={(current axis.north)},legend columns=-1}
    ]
\draw [black, densely dashed] (1.8,\pgfkeysvalueof{/pgfplots/ymin}) -- (1.8,\pgfkeysvalueof{/pgfplots/ymax});
\addplot [fill=gray!40!white] table [x=Nominal, y=Y] {\datatable};
\addplot [fill=red!60!white] table [x=Aged, y=Y] {\datatable};
\addlegendimage{blue,sharp plot}
\addlegendimage{red,sharp plot}
%\addlegendimage{black,densely dashed}
\legend{Time-zero delay,BTI-induced delay}
\end{axis}
\end{tikzpicture}%
}
%\hspace{2mm}
\pgfplotstableread{
Y Label Nominal Aged
1 Out7 1.320 0.4
2 Out6 1.672 0.1
3 Out5 1.320 0.4
4 Out4 1.672 0.1
5 Out3 1.320 0.4
6 Out2 1.672 0.1
7 Out1 1.320 0.4
8 Out0 1.5 0.25
}\datatable
%(Bars will be changed tomorrow)
\subfloat[\label{fig:path_delays_c}]{%
\begin{tikzpicture}
\begin{axis}[
    %ticks=none,
    xmajorticks=false,
    %axis x line=none,
    %axis y line=left,
    width=6cm,
    height=5cm,
    xbar stacked,
    %xmajorgrids,
    xmin=0,
    xmax=2.5,
    %xtick distance=0.5,
    enlarge y limits={0.1},
    bar width=3mm,
    ytick=data,
    yticklabels from table={\datatable}{Label}
    ]
\draw [black, densely dashed] (1.8,\pgfkeysvalueof{/pgfplots/ymin}) -- (1.8,\pgfkeysvalueof{/pgfplots/ymax});
\addplot [fill=gray!40!white] table [x=Nominal, y=Y] {\datatable};
\addplot [fill=red!60!white] table [x=Aged, y=Y] {\datatable};
\end{axis}
\end{tikzpicture}%
}
\caption{(a) Time-zero path delays of 3-to-8 pre-decoder. (b) Impact of aging. (c) Aging when rejuvenation workload ran together with main workload.}
\label{fig:path_delays}
\end{figure*}

The timing budget is set by the designer, and it is constant over the lifetime of the memory. As the pre-decoder suffers from aging, path delays of the decoder increases and more time required until pre-decoder outputs switch. Hence, less and less time is left for the slack. Eventually, when the slack time goes below zero, timing violations will occur.
%As discussed in the previous section, the slack is an important  reliability  metric  of  the  address  decoder. It  is  defined  as the .... Hence, the propagation delays of the pre-decoders are crucial for this metric. As the pre-decoder suffers from aging, its  path  delays  increase  and, thus, the  the  slack  decreases. Eventually,  when  the  slack  time  goes  below  zero,  timing violations will occur.

\cref{fig:path_delays_a} illustrates the delay of the longest path for each output in a 3-to-8 pre-decoder which are obtained by SPICE simulations. Activation delay of a path is the time required to pull the output to high. Likewise, deactivation delay corresponds to 1 to 0 output transition of a path. \cref{fig:path_delays_b} illustrates delay increase on decoder's paths due to aging. Since path delays of Out2, Out4, and Out6  reach beyond the timing budget, they cause delay faults in the memory. On the other hand, the other paths still have some slack time.

Transistors suffer from BTI aging only when they are under BTI stress. When a path's input switches, transistors in the path also switch. During activation or deactivation of an output, half of the transistors in the path will switch, thus half of them transitions from the BTI stress state (i.e., the biased condition) to the BTI relief state (i.e., the unbiased condition). The transistors that are in the relief state partially recover from the BTI aging effect. Thus, a path ages the most if workloads do not cause its input to switch~\cite{MJ16}, since a half of the transistors constantly stays at the BTI stress state and they have no chance to recover. Depending on the workload different addresses are selected and therefore different input combinations are applied to the decoder. Thus, a path's delay increase can be manipulated by workloads which run during memories operation~\cite{sw-based}.

%% file: 3-Proposed_Mitigation_Methodology.tex
\section{Proposed Mitigation Methodology}

The system that we consider as a case study in this work is composed of a CPU and a memory. The address output of the CPU is connected to the memory's address decoders. We analyze the larger (wordline) decoder and the least significant bits of the address signal are connected to the smaller (column) decoder. After the bits of the column decoder, 9 bits of the address signal are connected to the wordline decoder that has three 3-input pre-decoders. It should be noted that while the proposed methodology is explained using this specific example, it remains general and applicable for other memory decoder architectures. 

While the CPU executes its functional workload, memory operations (read/write) alters the address signal, hence the decoder's input. \cref{fig:path_delays_b} shows the path delays after aging while the CPU executes a given functional workload. It is clear that some paths in pre-decoder are longer compared to others. Furthermore, depending on the given workload, some paths have higher delay increase due to aging than others. As in \cite{sw-based}, we aim to mitigate aging by changing the workload that affects the memory. %However this workload change has execution overhead cost and it must be kept small or done in idle times.
This change is done by executing an dedicated small auxiliary workload periodically. We call such workload as a \emph{rejuvenation workload}, because it puts stressed transistors in long paths to the relief state to recover from the BTI aging. \cref{fig:path_delays_c} illustrates the resulting path delays after a rejuvenation workload is applied. Compared to \cref{fig:path_delays_b}, the longest delays of Out2, Out4, and Out6 are reduced, thus total pre-decoder's delay stays within the specified timing budget (shown as the vertical dashed line).

Our mitigation approach consists of three parts explained in following subsections. In III.A), we present the steps taken to calculate aging for a given workload. In III.B), three different rejuvenation workloads are presented with a method to generate them. In III.C), our mitigation scheme that combines both main and rejuvenation workloads together is explained.

\subsection{Aging Modeling and Assessment}

In order to analyze aging in the decoder we implement a flow that contains a high-level and a low-level setup. The high-level setup is responsible for generating a cycle-accurate \emph{memory trace} out of a given workload. The generated trace file must contain a cycle number and selected address pairs during the execution of the workload. The high-level setup can be realized with a RTL simulation of a CPU implementation or a cycle accurate instruction set simulation. the low-level setup has to be run at the transistor level and it is needed to obtain path delays in pre-decoders. The low level setup also includes the aging model.
\begin{figure}[ht]
\includegraphics[width=0.8\linewidth]{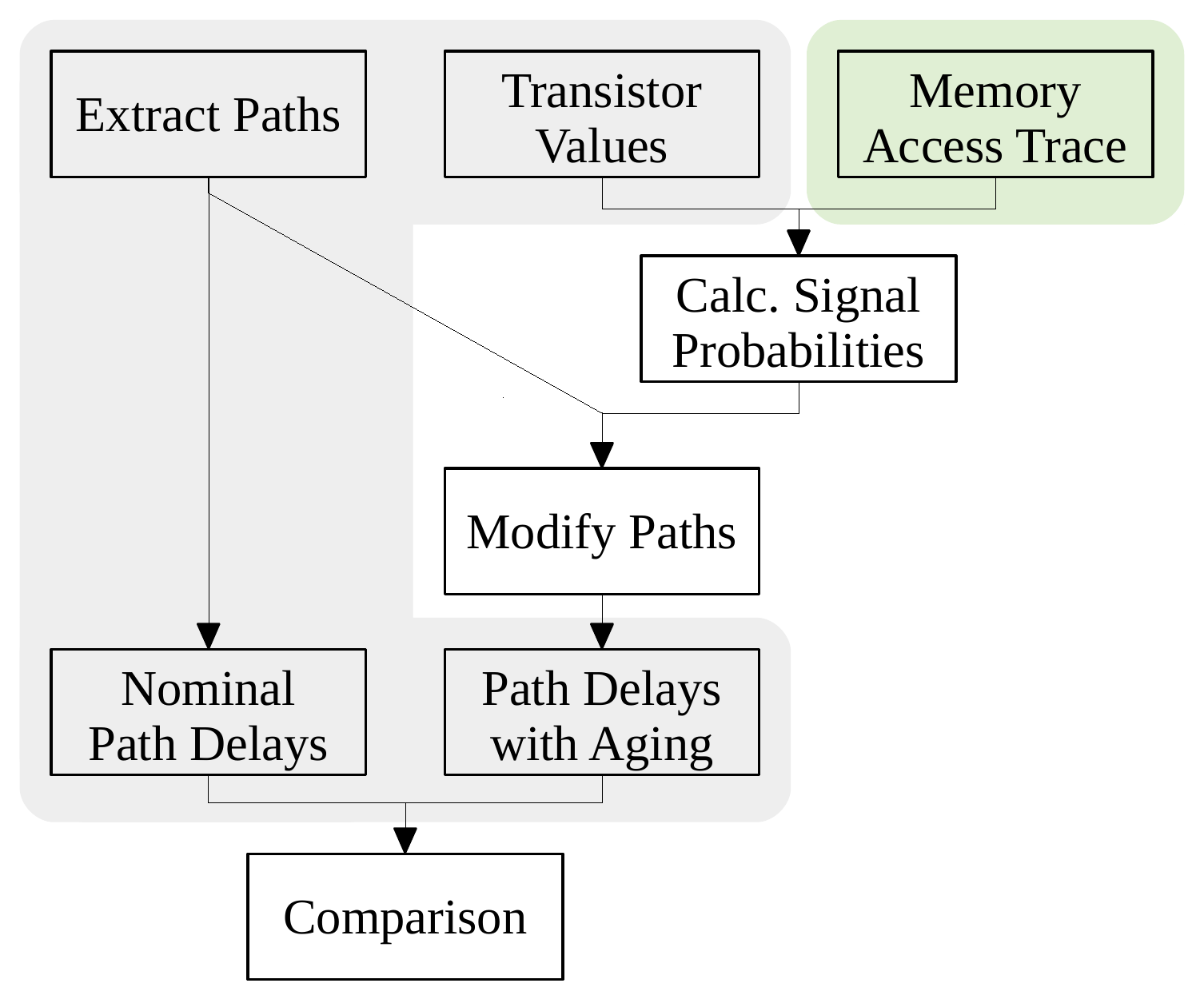}
\centering
\caption{The flow that is utilized to calculate aging in the decoder.}
\label{fig:flow}
\end{figure}

The flowchart in \cref{fig:flow} is explained below. The steps that are inside of the gray area requires a transistor-level simulation, and the green area corresponds to an RTL simulation. The rest of the steps are automated with scripts.

\begin{enumerate}[Step 1:]
\item Extract raw paths of the decoder from the decoder design. This step is executed once. (In particular, for the case-study design, 144 path files in transistor level are generated. Specifically, for each 3-to-8 pre-decoder, 24 activation and 24 deactivation paths exist.)
\item Identify and save transistor values in the decoder for each address input with transistor level simulation with transistor level simulation. This step is also executed once and transistor-value pairs are recorded for each address input.
\item Obtain nominal delay of each path with transistor level simulation.
\item Obtain memory access traces by running RTL simulation.
\item Calculate signal probabilities for each transistor.
\item Add average signal values of transistors to each path file.
\item Obtain delay of each path after aging by running transistor level simulation with aging model. Note, since the critical path may change after aging, we simulate all paths in this step.
\item Calculate aging percentage using the results from Step 4 and 7. 
\end{enumerate}

\begin{figure*}[ht!]
\centering
\pgfplotstableread{
X   Label Nominal     min   dwrej    drej    urej   norej
1  aescbc     100 105.504 112.994 114.112 112.674 114.209
2  conv2d     100 105.504 113.365 115.916 113.190 118.806
3 fdctfst     100 105.504 114.576 114.577 114.702 115.303
4     fft     100 105.504 119.241 119.412 120.194 129.834
5     fir     100 105.504 119.690 119.909 120.875 134.110
6     ipm     100 105.504 112.722 114.212 112.786 114.331
7  keccak     100 105.504 114.081 115.100 113.453 115.212
8     sha     100 105.504 115.274 116.237 114.656 118.856
9    avg.     100 105.504 115.243 116.184 115.316 120.083
}\tableANDAND
\pgfplotstableread{
X   Label Nominal     min   dwrej    drej    urej   norej
1  aescbc     100 108.761 121.420 121.421 123.706 125.919
2  conv2d     100 108.761 124.899 126.079 124.244 130.269
3 fdctfst     100 108.761 121.203 123.958 123.323 125.748
4     fft     100 108.761 121.601 123.800 126.991 129.690
5     fir     100 108.761 123.226 124.128 122.121 128.805
6     ipm     100 108.761 121.306 122.020 123.802 126.018
7  keccak     100 108.761 122.506 122.499 124.379 126.843
8     sha     100 108.761 125.659 126.090 125.599 130.317
9    avg.     100 108.761 122.728 123.749 124.271 127.951
}\tableNANDNOR
\subfloat[AND-AND\label{fig:chart1}]{%
%{todo: subtract 100 from y axis}\sub
\begin{tikzpicture}
	\begin{axis}[
	%width=8.2cm,
    %height=7cm,
	%ybar stacked,
	ybar,
	bar shift=0,
	ymin=97,
	ymax=135,
	scale=0.8,
	ymajorgrids,
	ytick distance=5,
	xtick=data,
	xtick={1,...,9},
	xticklabels from table={\tableANDAND}{Label},
	legend style={anchor=north east,legend columns=2,font=\fontsize{7pt}{1pt}\selectfont},
	%symbolic x coords={aescbc,conv2d,fdctfst,fft,fir,ipm,keccak,sha},
	x tick label style={
       xshift=2mm,
       yshift=-1mm,
       rotate=45,
       anchor=east,
    },
    ylabel={Aging},
    yticklabel=\pgfmathparse{\tick}\pgfmathprintnumber{\pgfmathresult}\%\
	]
	\addplot [fill=red!60!white] table [x=X, y=norej] {\tableANDAND};
	\addplot [fill=orange!60!white] table [x=X, y=urej] {\tableANDAND};
	\addplot [fill=yellow!60!white] table [x=X, y=drej] {\tableANDAND};
	\addplot [fill=brown!40!white] table [x=X, y=dwrej] {\tableANDAND};
	\addplot [fill=white,postaction={pattern=north west lines}] table [x=X, y=min] {\tableANDAND};
	\addplot [fill=gray!40!white] table [x=X, y=Nominal] {\tableANDAND};
	\legend{No Rej,Uni.,Des., D\&W,Min.,Nom.}
	\end{axis}
\end{tikzpicture}%
}
\hspace{5mm}
\subfloat[NAND-NOR\label{fig:chart2}]{%
\begin{tikzpicture}
	\begin{axis}[
	%width=8.2cm,
    %height=7cm,
	%ybar stacked,
	ybar,
	bar shift=0,
	ymin=97,
	ymax=135,
	scale=0.8,
	ymajorgrids,
	ytick distance=5,
	xtick=data,
	xtick={1,...,9},
	xticklabels from table={\tableNANDNOR}{Label},
	%symbolic x coords={aescbc,conv2d,fdctfst,fft,fir,ipm,keccak,sha},
	x tick label style={
       xshift=2mm,
       yshift=-1mm,
       rotate=45,
       anchor=east,
    },
     ylabel={Aging},
    yticklabel=\pgfmathparse{\tick}\pgfmathprintnumber{\pgfmathresult}\%\
	]
	\addplot [fill=red!60!white] table [x=X, y=norej] {\tableNANDNOR};
	\addplot [fill=orange!60!white] table [x=X, y=urej] {\tableNANDNOR};
	\addplot [fill=yellow!60!white] table [x=X, y=drej] {\tableNANDNOR};
	\addplot [fill=brown!40!white] table [x=X, y=dwrej] {\tableNANDNOR};
	\addplot [fill=white,postaction={pattern=north west lines}] table [x=X, y=min] {\tableNANDNOR};
	\addplot [fill=gray!40!white] table [x=X, y=Nominal] {\tableNANDNOR};
	\end{axis}
\end{tikzpicture}%
}
\caption{Comparison of eight workloads with different mitigation strategies (at 1\% execution overhead and 3 years of aging).}
\label{fig:chart12}
\end{figure*}
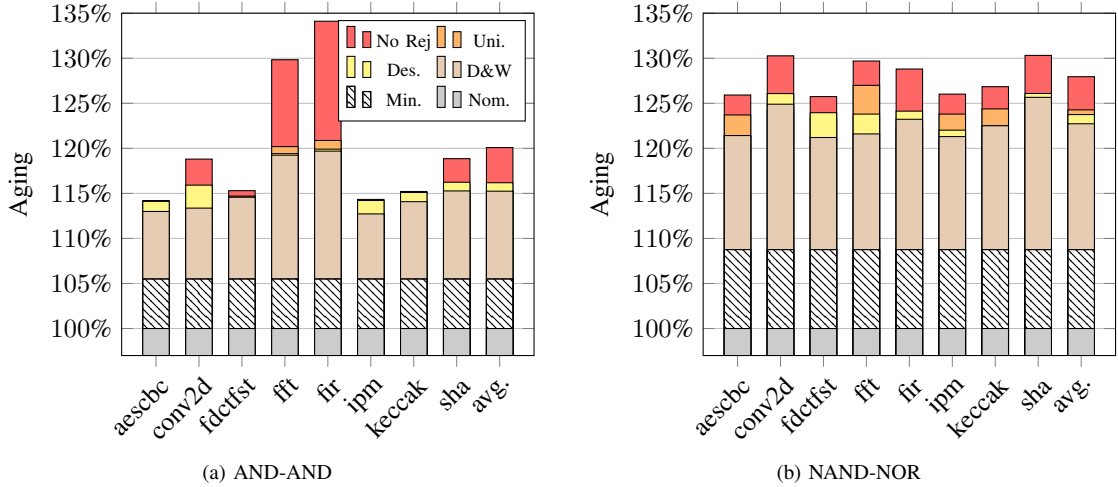
\subsection{Rejuvenation Workload Generation}

Here, we propose several rejuvenation workloads and provide a method to generate them.

\begin{itemize}
\item \textbf{Universal:} This rejuvenation workload sequentially selects all addresses for equal amount of clock cycles as in \cite{sw-based}. This is the simplest workload out of the three, and can be generated without any analysis of decoder design or memory access trace.

\item \textbf{Design-Aware:} This rejuvenation workload takes path delays into account, and selects some of the addresses for more cycles than others to balance path delays. \cref{fig:path_delays_c} shows the path delays of a pre-decoder after design aware workload run for the specified period (e.g., three years in the case study). To generate this workload several iterations of aging assessment has to be made. In the given example, 3-bit pre-decoder inputs can be controlled by 8 addresses. Initially all 8 addresses are selected by equal amount of time. After the first run, we take a note of the longest path and its input value. Then for the second run we increase cycles for address that corresponds to that input value. This process is repeated until all of the long paths reach the same amount of delay and no single path has a higher delay than others. At the end of this process we obtain n \emph{weights} (constant values) for each of the n addresses. To achieve lowest aging, selection ratio of addresses (in cycles) must match the weights which are obtained for the design. It is important that, this rejuvenation workload is generated once for a given decoder design.

\item \textbf{Design-\&-Workload-Aware:} This rejuvenation workload uses the weights that are calculated for Design-Aware workload and combines it with the main functional workload's memory trace. Based on its memory trace, we calculate signal probability of each pre-decoder input, and generate a rejuvenation workload that balances it. For instance, most workloads use the last addresses of the decoder more frequently as those corresponds to the stack portion of the memory. We generate a workload such that those highly utilized addresses are selected less while the rejuvenation workload is running. This process yields the best rejuvenation for a given time overhead. However, it requires the knowledge of both the design and the main workload ahead of time.
\end{itemize}

\subsection{On Aapplying the Rejuvenation Workloads}

After deciding and generating the rejuvenation workload, it must be mixed to the main functional workload. In our mitigation scheme, the rejuvenation workload is added to a interrupt routine to be applied periodically. The routine is called with a timer interrupt and by adjusting its count limit, the routine can be called at the desired frequency and execution overhead. The routine can be placed in either \emph{crt0} file, so that it can be compiled without modifying the main code, or it can be attached to the main functional workload.

%% file: 4-Experimental_Results.tex
\section{Experimental Results}
\subsection{Experimental Setup}

Our experimental setup consists of two parts. The high-level part, which is used to generate memory access traces, is implemented with some modifications to Pulpino~\cite{traber2016pulpino} open-source single core microcontroller project. The RISCY CPU in the project configured to implement RISC-V RV32F instruction set. We increased the observability of the memory, since we are also interested in memory state when the CPU is at sleep state. We used the benchmarks in the design repository, some of them originating from the MiBench library ~\cite{guthaus2001mibench}. The unrelated benchmark statistics code and UART messages were removed to observe the workload only. The interrupt routine was implemented in C and the rejuvenation workload was implemented as an assembly code. The interrupt period is adjusted based on the length of both the rejuvenation and the main workloads to achieve desired execution overhead. At the end of RTL simulation, we obtain a memory trace file that contains memory state during the execution of the workload, i.e. read/write operations, the selected address and the corresponding clock cycle.

As mentioned in Section II, we consider two 9-to-512 decoder designs using 22nm PTM technology and the aging model~\cite{rzepa2018comphy} %(28nm Comphy??)
on the low-level setup. We used Spectre~\cite{kundert2006designer} to do SPICE simulations for extracting paths of the pre-decoders, obtaining transistor values for a given address input and at the end of the flow, to obtain path delays with or without aging. We also have a Python script to automate the flow. It takes memory trace as an input, calculates signal probabilities, adds the signal probability values to the transistors on a the paths and runs SPICE simulations for each path. We assume that simulations with the aging model are run at 125°C and a nominal supply voltage of 0.95V. The duration of the aging is 3 years except the last experiment, where simulations are run for 1 to 10 years with one year increments.

The assembly code for the universal rejuvenation workload is implemented manually. However, they are generated with a script for the other methods.

\subsection{Performed Experiments}

Using the setup above, the following experiments are performed:

\subsubsection{Dependency on Functional Workload}

We first investigate how much the functional workload affects the mitigation potential to see the best and the worst cases. On the later experiments, we use averages of all workloads or a single one. \cref{fig:chart12} shows all the functional/rejuvenation workload combinations for 3 years of aging at 1\% execution overhead.

For AND-AND design there is much larger deviation from the average value. The main difference between the two designs is on the AND-AND design, lowest outputs of the pre-decoder has longer delay, and higher ones have lower delays. This is the opposite on the NAND-NOR design. Therefore, to mitigate aging with the Design-Aware rejuvenation, we allocate more clock cycles for the high address range of the memory with the AND-AND design. The \emph{fir} and the \emph{fft} workloads heavily utilize low address ranges of the memory but they rarely use the stack for function calls, hence the 3rd pre-decoder that connected to most significant three bits of the address signal stays at "000" input for 95\% of the time. Due to this imbalance, these two benchmarks have the highest aging on AND-AND design and the \emph{aescbc} has already 60\% less aging compared to \emph{fir}. By adding 1\% rejuvenation overhead we reduce this difference to  $\sim$20\%.

% fig reduction
\begin{figure}[h!]
%\centering
\begin{tikzpicture}
\begin{axis}[
    ybar=.2cm,
    ymin=0,
    x = 3cm,
    enlarge x limits={abs=1.2cm},
    scale=0.8,
    legend style={anchor=north east,legend columns=1,font=\fontsize{8pt}{1pt}\selectfont},
    %legend cell align={left},
    ylabel={Aging reduction},
    ymajorgrids,
    ytick distance=5,
    symbolic x coords={AND-AND,NAND-NOR},
    xtick=data,
    nodes near coords={\pgfmathprintnumber\pgfplotspointmeta\%},
    nodes near coords align={vertical},
    yticklabel=\pgfmathparse{\tick}\pgfmathprintnumber{\pgfmathresult}\%\
    ]
\addplot [fill=orange!60!white] plot [error bars/.cd, y dir = both, y explicit] coordinates {(AND-AND,23.7)+=(0,15.100)-=(0,19.769) (NAND-NOR,13.1)+=(0,10.107)-=(0,4.585)};
\addplot [fill=yellow!60!white] plot [error bars/.cd, y dir = both, y explicit] coordinates {(AND-AND,19.4)+=(0,22.233)-=(0,18.718) (NAND-NOR,15.0)+=(0,4.840)-=(0,8.049)};
\addplot [fill=brown!40!white] plot [error bars/.cd, y dir = both, y explicit] coordinates {(AND-AND,24.1)+=(0,18.175)-=(0,19.346) (NAND-NOR,18.7)+=(0,8.547)-=(0,3.337)};
\legend{Universal,Design Aware,D\&W Aware}
\end{axis}
\end{tikzpicture}
\caption{Aging reduction of three rejuvenation workloads compared to no rejuvenation after 3 years.}
\label{fig:reduction}
\end{figure}
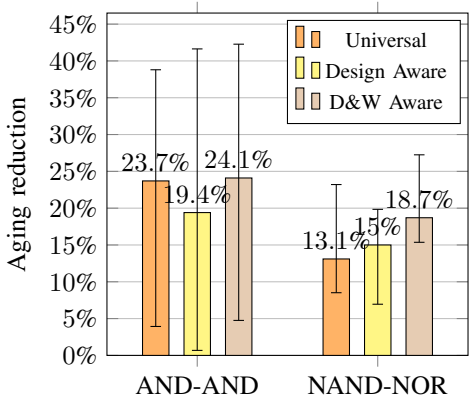

\subsubsection{Dependency on Rejuvenation Workload}

Here, we look at the same data as the previous one, but we compare the impact of rejuvenation workloads.

\cref{fig:reduction} shows the average aging reduction for three rejuvenation workloads. The error bars show the results with best and worst case considering the eight main workloads. We achieved the best results with the Design-\&-Workload-Aware rejuvenation, which yield to a greater benefit over the others on the average as well as best/worst cases. It has reduced aging by 42\% in the best case and by 5\% in the worst one. We did not observe a clear advantage with Design-Aware workload over the Universal.
\subsubsection{Saturation of The Rejuvenation Effect}

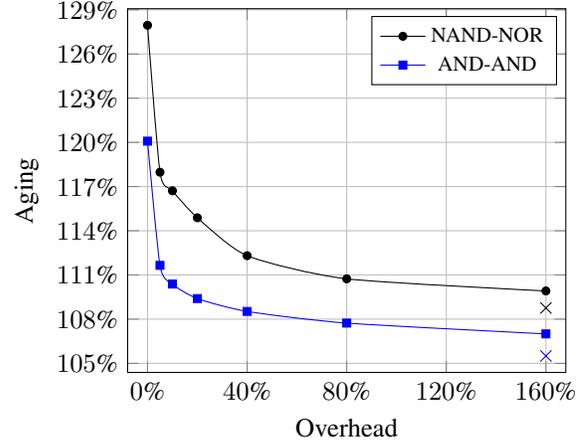
\begin{figure}[t!]
\centering
\pgfplotstableread{
Overhead ANDANDdwrej NANDNORdwrej
       0     120.083      127.951
       5     111.652      117.974
      10     110.382      116.715
      20     109.398      114.883
      40     108.525      112.309
      80     107.733      110.737
     160     107.000      109.915
}\tableOverhead
\begin{tikzpicture}
	\begin{axis}[
	    xtick={0,40,...,160},
	    xticklabels from table={\tableOverhead}{Overhead},
	    ytick distance=3,
	    scale=0.85,
	    %ymajorgrids,
	    grid=major,
		xlabel=Overhead,
		ylabel=Aging,
		%xmode=log,
		log basis x=2,
		enlarge x limits={0.05},
		enlarge y limits={0.05},
		legend style={anchor=north east,legend columns=1,font=\fontsize{8pt}{1pt}\selectfont},
		xticklabel=\pgfmathparse{\tick}\pgfmathprintnumber{\pgfmathresult}\%,
		yticklabel=\pgfmathparse{\tick}\pgfmathprintnumber{\pgfmathresult}\%\
	]
	\addplot +[smooth,tension=0.4,samples=1000,mark size=1.5,mark options={black}]  [draw=black] table [x=Overhead, y=NANDNORdwrej] {\tableOverhead};
	\addplot +[smooth,tension=0.4,samples=1000,mark size=1.5,mark options={blue}] [draw=blue] table [x=Overhead, y=ANDANDdwrej] {\tableOverhead};
	\addplot [draw=blue,fill=blue,mark=x,mark size=3] coordinates{(160, 105.504)};
	\addplot [draw=black,fill=black,mark=x,mark size=3] coordinates{(160, 108.761)};
	%\addplot [draw=blue,fill=blue,mark=triangle*,mark size=2] coordinates{(160, 107.829)}; %universal rej
	\legend{NAND-NOR,AND-AND}
	\end{axis}
\end{tikzpicture}
\caption{Average aging in 3 years vs. execution overhead. The cross marks show the minimum aging possible.}
\label{fig:overhead}
\end{figure}

In this experiment we investigate the saturation of the rejuvenation effect as we increase the execution overhead. The \cref{fig:overhead} shows the average aging considering all main workloads versus the execution overhead. It is clear that we get the highest amount of benefit with the first few percentages of the rejuvenation execution overhead. This expected as the transistors suffer from extreme static BTI effect when the input signal probability is close to 1.0 or 0.0, but the effect quickly decreases (see \cref{fig:overhead}). The crosses mark the lowest possible aging (when running only the Design-Aware workload).
% fig overhead
% fig years
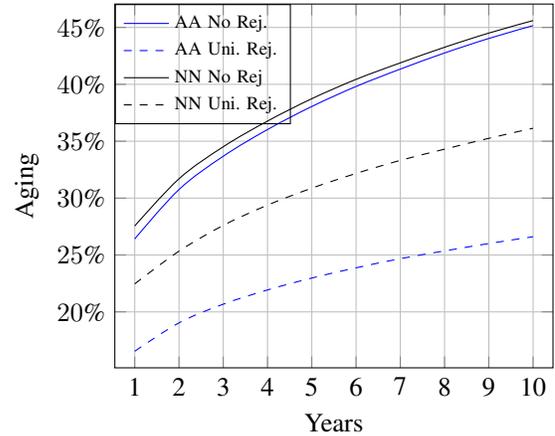
\begin{figure}[t!]
\centering
\pgfplotstableread{
Years ANDANDnorej ANDANDurej NANDNORnorej NANDNORurej
    1      26.409     16.537       27.549      22.462
    2      30.755     19.036       31.697      25.335
    3      33.678     20.676       34.513      27.601
    4      36.026     21.927       36.770      29.383
    5      38.051     22.974       38.736      30.873
    6      39.811     23.871       40.436      32.166
    7      41.339     24.677       41.885      33.316
    8      42.754     25.357       43.258      34.300
    9      44.032     26.006       44.505      35.255
   10      45.162     26.606       45.601      36.141
}\tableYears
\begin{tikzpicture}
	\begin{axis}[
	    xtick={1,...,10},
	    xticklabels from table={\tableYears}{Years},
	    ytick distance=5,
	    scale=0.85,
	    %ymajorgrids,
	    grid=major,
		xlabel=Years,
		ylabel=Aging,
		%xmode=log,
		%log basis x=2,
		%ymin=0,
		%enlargelimits=false,
		enlarge x limits={0.05},
		enlarge y limits={0.05},
		legend style={at={(0,1)},anchor=north west,legend columns=1,fill=none, font=\fontsize{7pt}{1pt}\selectfont},
		legend cell align={left},
		yticklabel=\pgfmathparse{\tick}\pgfmathprintnumber{\pgfmathresult}\%\
	]
	\addplot +[smooth,tension=0.6,samples=1000,no marks] [draw=blue] table [x=Years, y=ANDANDnorej] {\tableYears};
	\addplot +[smooth,tension=0.6,samples=1000,no marks] [draw=blue,dashed] table [x=Years, y=ANDANDurej] {\tableYears};
	\addplot +[smooth,tension=0.6,samples=1000,no marks] [draw=black] table [x=Years, y=NANDNORnorej] {\tableYears};
	\addplot +[smooth,tension=0.6,samples=1000,no marks] [draw=black,dashed] table [x=Years, y=NANDNORurej] {\tableYears};

	%\addplot [draw=blue,fill=blue,mark=x,mark size=3] coordinates{(160, 105.504)};
	%\addplot [draw=red,fill=red,mark=x,mark size=3] coordinates{(160, 108.761)};
	\legend{AA No Rej.,AA Uni. Rej.,NN No Rej,NN Uni. Rej.}
	\end{axis}
\end{tikzpicture}
\caption{Aging of the workload fir with and without the universal rejuvenation at 1\% overhead, simulated for 1 to 10 years range.}
\label{fig:years}
\end{figure}

\subsubsection{Dependency on Target Lifetime}

Lastly, we performed an experiment to estimate the amount of the system's lifetime extension by our mitigation method. We ran cases from 1 year of aging to 10 years with 1 year increments with and without Universal rejuvenation. The mitigation scheme applied assuming \emph{fir} as the functional workload with 1\% overhead. The \cref{fig:years} shows the amount of aging while duration of aging increased. Blue and black lines represents AND-AND and NAND-NOR decoder design respectively. The results demonstrate that, for AND-AND design, the same amount of aging accumulated after 9 years of aging with mitigation and 1 year of aging without mitigation. It is possible to observe a similar benefit with NAND-NOR design, where we see the same amount of aging at 9 and 3 years with and without mitigation respectively. the actual extended lifetime strongly depends on the initially allocated slack, but generally it may be up to several times as in the discussed case-study.

%% file: 5-Discussion.tex
\section{Discussion}
In this section, we make following points on the performed experiments.

\begin{enumerate}[A.]
%\subsection{Improved Reliability} 
\item \emph{Improved Reliability}
Our experimental results show that we achieved 21\% aging reduction on average, up to 42\% in some cases. We also reflected this on the life time of decoder and showed that it is possible to extend duration of its reliable operation by three times or more.

\item\emph{Cost of the scheme}
Since our mitigation scheme implemented as a software addition to the original code, we do not need a new hardware design, and therefore there is no area or path delay overhead. However, the scheme allocates CPU time and has an execution time overhead. In our experiments, we set this overhead to only 1\% and achieved significant benefits. In addition, we showed that it is possible to gain most of the rejuvenation benefit in first few percentages of execution overhead. Due to the execution overhead, our scheme has also a power overhead.

Our scheme is also transparent from the software point of view, and easy to integrate. Although a new compilation of the software necessary, it is possible to add it to the compiler and keep the existing programs the same.

\item \emph{Choosing the rejuvenation workload}
In our experiments the NAND-NOR decoder design reached 43\% more aging reduction with Design and Workload Aware rejuvenation workload compared to the Universal one. However, we did not observed this on the other decoder design. This is mostly depends on the pre-decoder paths, since the AND-AND had long paths connected to lowest bits, and the opposite with the NAND-NOR design.

Choosing the rejuvenation workload depends on the system and also the workloads that the CPU needs to execute. We observed that most workloads use first and last part of the memory, and they can benefit from a workload aware rejuvenation. In most systems however, it will not be possible to use a workload aware method if an operating system is present, or on a multi-core system. In such systems, Design-Aware or Universal methods may be applied.

\item \emph{Potential Improvements}
In our experiments, we assume that the CPU busy 100\% of the time. In reality, there can be idle periods that can be used to run our mitigation scheme. This may also reduce or eliminate the execution overhead. We did not explore this, since it is too dependant on the system, but presented the data so that the benefit can be calculated. Our results showed that we can gain almost half of maximum possible aging reduction with a 5\% of overhead and it quickly saturates as we increase the overhead percentage.
\end{enumerate}

%% file: 6-Conclusion.tex
\section{Conclusion}

In this work, we presented a low-cost aging mitigation scheme, which can be applied to existing hardware to mitigate aging on memory address decoder logic. We mitigate the BTI effect on critical transistors by applying a rejuvenation workload to the memory. Such an auxiliary workload is executed periodically to rejuvenate transistors that are located on critical paths of the address decoder. Second, we analyze workloads’ efficiency to optimize the mitigation scheme. Experimental results performed with realistic benchmarks demonstrate several-times lifetime extension with a negligible execution overhead.